\documentclass[showpacs,aps,prl,twocolumn,epsfig]{revtex4}
\usepackage{graphicx}
\usepackage{amsfonts}

\def\eps{\varepsilon}

\begin{document}
\title{Dynamical stabilization of solitons in cubic-quintic nonlinear Schr\"odinger
model}
\author{Fatkhulla Kh. Abdullaev\dag \, and Josselin Garnier\ddag\footnote[7]{Corresponding author (garnier@math.jussieu.fr)}
}
\address{\dag\
Dipartimento di Fisica "E.R. Caianiello", Universit\'a di Salerno, 84081 Baronissi (SA), Italy\\
\ddag\ Laboratoire de Probabilit\'es et Mod\`eles Al\'eatoires
\& Laboratoire Jacques-Louis Lions,
Universit{\'e} Paris VII,
2 Place Jussieu, 75251 Paris Cedex 5,
France}

\begin{abstract}
We consider the existence of a dynamically stable  soliton
in the one-dimensional cubic-quintic nonlinear Schr\"odinger
model with strong cubic nonlinearity management for periodic and random modulations.
 We show that the predictions of the
averaged cubic-quintic NLS equation and modified variational
approach for the arrest of collapse  coincide.
The analytical results are confirmed by
 numerical
simulations of one-dimensional cubic-quintic NLS equation  with
rapidly and strongly varying cubic nonlinearity coefficient.
\end{abstract}
\maketitle

\newpage


Collapse phenomena are observed in many areas of physics:
self-focusing of intense laser beams, Langmuir waves in plasma,
collapse of the Bose-Einstein condensates (BECs)  with attractive interactions
etc.
The nonlinear Schr\"odinger equation (NLSE) with cubic nonlinearity
used to describe these systems has stable solutions in the one-dimensional (1D) case,
when the dispersion and nonlinearity effects can effectively balance
each other. In two and three dimensions the focusing nonlinearity
overcomes the dispersion and the blow-up phenomenon occurs \cite{Sulem}.

Few mechanisms for the arrest of collapse have been suggested.
Among them can be mentioned the dispersion \cite{Zhar,Abd1} and
nonlinearity management methods
\cite{Berge2,Towers,Abd2,Saito1,Perez}.  The analysis based on the
variational approach, method of moments  and numerical simulations
showed that the nonlinearity management method is effective to
suppress collapse in the scalar and vector 2D NLSE with
focusing cubic nonlinearity. For the 3D cubic NLSE with
nonlinearity management theoretical predictions and numerical results
do not bring a clear and definitive picture \cite{Saito2,Adhikari},
so more analytical and numerical work is necessary.

In this Rapid Communication we investigate the phenomenon of arrest
of collapse by using the strong cubic nonlinearity management scheme in the
1D cubic-quintic (CQ) NLSE.
The CQ NLSE with nonlinearity management presents practical
interest since it appears in
many branches of physics such as  nonlinear optics and BEC.
In nonlinear optics it describes the propagation of pulses in double-doped optical
fibers \cite{DeAngelis}, in BEC it models the condensate
with two and three body interactions \cite{Abd3,Meystre}.
In optical fibers periodic variation of the nonlinearity can be achieved by
varying the type of dopants along the fiber. In BEC the variation of the atomic
scattering length by the Feshbach resonance technique leads to the oscillations of the
mean field cubic nonlinearity \cite{Inouye}.

The CQ NLSE when the cubic term is equal to zero is the critical
quintic NLSE. The quintic Townes soliton is an unstable solution
of the quintic NLSE \cite{Sulem}. In this work we consider the
configuration with the rapid and strong periodic modulation in
time  of cubic nonlinear interaction. This type of modulations
corresponds to the management applied to the nonlinearity with a
lower power and has never  been studied. We first apply a
variational approach to the averaged NLSE. The averaged equation
for the 1D cubic NLSE in the case of strong nonlinearity
management has been derived in \cite{Kevrekidis1,Pelinovsky}, but
the presence of the quintic term dramatically changes the picture
as this term would normally lead to collapse. We also propose a
modified variational approach for the managed CQ NLSE where the
ansatz is designed to take into account the fast self-phase
modulation due to the cubic nonlinearity management. These two
approaches predict the stabilization of the quintic Townes
soliton.

We consider the CQ NLSE
\begin{equation}
\label{eq:cq}
i u_t + u_{xx} + \gamma(t) |u|^2 u + \chi |u|^4 u =0 \, ,
\end{equation}
with an attractive quintic nonlinearity $\chi >0$.
The time-varying cubic coefficient $\gamma(t)$ possesses an average value
$\gamma_0$ and a fast varying part $\gamma_1$
\begin{equation}
\gamma(t ) = \gamma_0 + \frac{1}{\eps} \gamma_1(\frac{t}{\eps} ) \, ,
\end{equation}
where $\eps \ll 1$ corresponding to strong and rapid management.
Here $\gamma_1$ can be either a periodic function or
a stationary random function.
We first address the case of a periodic management $\gamma_1(\tau+1)=\gamma_1(\tau)$
and $\int_0^1 \gamma_1 (\tau) d \tau =0$.
We introduce $\Gamma_1(\tau) = \int_0^\tau \gamma_1(s) ds$
which is also  a zero-mean periodic  function.
Following the same procedure  as in \cite{Pelinovsky},
we can average the CQ NLSE over fast variations
and show that the solution takes the form
\begin{eqnarray*}
u(t,x)=\left[ w(t,x) + \eps w_1(t,\frac{t}{\eps},x) + \cdots \right] \\
\times \exp \left[ i   \Gamma_1(\frac{t}{\eps}) |w(t,x)|^2 \right] \, ,
\end{eqnarray*}
where $w$ is solution of the averaged CQ NLSE
\pagebreak
\begin{eqnarray}
\nonumber
\label{eq:cqave}
&& i w_t + w_{xx} + \gamma_0 |w|^2 w + \chi |w|^4 w  \\
\label{eq:cqave}
&& + \sigma^2
\left\{ [(|w|^2)_x]^2+2 |w|^2 (|w|^2)_{xx} \right\} w=0 \, ,
\end{eqnarray}
and $\sigma^2 = \int_0^1 \Gamma_1(s)^2 ds$.
The averaged CQ NLSE has a Hamiltonian form, with the Hamiltonian
\begin{equation}
\label{eq:hamave} H = \int (|w_{x}|^2 -\frac{ \gamma_0 }{2} |w|^4
- \frac{ \chi }{3}|w|^6 + \sigma^2 [(|w|^2)_x]^2 |w|^2 ) dx \, .
\end{equation}
We first prove that the solution of Eq.~(\ref{eq:cqave}) cannot collapse
because its supremum norm can be a priori bounded.
This proof is essentially based on the Sobolev inequality $\| f \|_\infty^2 \leq
C \|Êf \|_2  \|Êf_x \|_2$ where we denote $\|f\|_p =
\left[ \int |f(x)|^p dx \right]^{1/p}$ for $p \in (1,\infty)$ and
$\|f\|_\infty $ is the (essential) supremum of $|f|$.
We first apply this inequality with $f=w$:
\begin{equation}
\label{eq:sob1}
\| w \|_4^4 \leq C \| w \|_\infty^2  \| w \|_2^2 \leq C \|Êw \|_2^3  \|Êw_x \|_2
\leq C  \|Êw_x \|_2 \, ,
\end{equation}
where $C$ stands for a constant that may change from line to line
and we have used the fact that $\| w \|_2$ is constant.
Next we apply the Sobolev inequality with $f=v := |w|^3$:
\begin{eqnarray*}
\| w \|_6^6 &\leq& \|w\|_\infty^4 \|w\|_2^2 = \| v \|_\infty^{4/3}  \|w\|_2^2
\leq C  \|Êv \|_2^{2/3}  \|Êv_x \|_2^{2/3} \\
&\leq&  C\delta  \|Êv \|_2^{4/3} + C \delta^{-1} \|Êv_x \|_2^{4/3} \, .
\end{eqnarray*}
The last estimate holds true for any $\delta>0$, so by choosing
$\delta = \| v \|_2^{2/3} /(2C) $ and noting that
$\|Êv \|_2^2 = \| w \|_6^6$,
we get
\begin{equation}
\label{eq:sob2}
\| w \|_6^6 \leq C  \| v_x \|_2 \, .
\end{equation}
Substituting (\ref{eq:sob1}-\ref{eq:sob2}) into (\ref{eq:hamave})
and using the fact that  $\int [(|w|^2)_x]^2 |w|^2 dx = (4/9)  \| v_x \|_2$,
we can finally write
$$
H \geq  \|Êw_x \|_2^2 - \gamma_0 C  \|Êw_x \|_2 + (4/9) \sigma^2  \| v_x \|_2 ^2
- \chi C  \| v_x \|_2 \, ,
$$
which shows that $\|Êw_x \|$ and $\| v_x \|$, and thus $\|w\|_\infty$
are uniformly bounded, which prevents the solution of the averaged
CQ NLSE  from collapsing.
However, this result does not show that the solution of Eq.~(\ref{eq:cqave})
does not collapse for fixed $\eps$.

The next step consists in applying the
variational approach to the averaged CQ NLSE  \cite{anderson83,malomed02}.
The variational ansatz for the solution is chosen as
the chirped function
\begin{equation}
w(t,x)  = A(t) Q\left( \frac{x}{a(t)}\right) \exp \left[ i b(t) x^2+ i \phi(t) \right] \, ,
\end{equation}
with a given shape $Q$.
Following the standard procedure,
we substitute the ansatz into the Lagrangian density generating Eq.~(\ref{eq:cqave})
and calculate the effective Lagrangian density in terms of $A$, $a$, $b$, $\phi$ and their time derivatives. The evolution equations
for the parameters of the ansatz are then derived from the effective Lagrangian by using the
corresponding Euler-Lagrange equations. In particular this approach yields a closed-form ordinary
differential equation (ODE) for the width $a$:
 \begin{equation}
 \label{eq:odeave}
 a_{tt} = - U_{av}' (a) \, ,
 \end{equation}
where the effective potential is
\begin{equation}
U_{av}(a) = \frac{ 6 D_1- 2 D_3 \chi N^2}{3a^2}
- \frac{D_2 \gamma_0 N }{a} + \frac{8 D_4 \sigma^2 N^2}{a^4} \, ,
\end{equation}
the total mass is $N=\int |u|^2 dx$ and the effective parameters are
\begin{eqnarray*}
D_1 &=& \frac{ \int Q'(x)^2 dx}{\int x^2 Q^2(x) dx} \, ,\\
D_2 &=& \frac{ \int Q(x)^4 dx}{\left( \int Q^2(x) dx\right) \left( \int x^2 Q^2(x) dx\right)} \, ,\\
D_3 &=& \frac{ \int Q(x)^6 dx}{\left( \int Q^2(x) dx\right)^2 \left( \int x^2 Q^2(x) dx\right)} \, , \\
D_4 &=& \frac{ \int Q(x)^4 Q'(x)^2 dx}{\left( \int Q^2(x) dx\right)^2 \left( \int x^2 Q^2(x) dx\right)} \, .
\end{eqnarray*}

Another approach is possible for the CQ NLSE
with cubic nonlinearity management (\ref{eq:cq}).
It consists in applying a specific variational approach
designed to capture the fast self-phase modulation induced
by the fast nonlinearity management.
The variational ansatz is sought in the form
\begin{eqnarray}
\nonumber
u(t,x)  &=& A(t) Q\left( \frac{x}{a(t)}\right)
\exp\left[ i b(t) x^2+ i \phi(t) \right] \\
&& \times \exp\left[ i \Gamma_1(\frac{t}{\eps}) A^2(t) Q^2(\frac{x}{a(t)})  \right] \, .
\label{eq:ansatz}
\end{eqnarray}
Substituting this ansatz into the Lagrangian density generating Eq.~(\ref{eq:cq})
we get the system of ODEs
\begin{eqnarray}
\label{eq:odea}
a_t &=& 4ab - \frac{D_2 N }{a^2} \Gamma_1( \frac{t}{\eps}) \, , \\
\nonumber
b_t &=& \frac{D_1}{a^4} - 4 b^2 -\frac{D_2 \gamma_0 N}{4 a^3} -
\frac{D_3 \chi N^2}{3 a^4} - \frac{D_2 N b}{a^3} \Gamma_1(\frac{t}{\eps}) \\
\label{eq:odeb}
&&
+ \frac{8 D_4 N^2}{a^6} \Gamma_1^2 (\frac{t}{\eps}) \, .
\end{eqnarray}
Next we perform an averaging of this system of ODEs, which yields
exactly Eq.~(\ref{eq:odeave}).
This analysis shows that the two different approaches yield the same effective
system for the soliton parameters, which strengthens its validity.

We now focus our attention to the particular case $\gamma_0=0$.
We first choose the shape function $Q$
corresponding to the quintic Townes soliton $
Q(x) = {1}/{\sqrt{\cosh(x)}}$,
which gives the values $D_1=1/(2\pi^2)$,
$D_2=8/\pi^4$, $D_3 = 2/\pi^4$, and $D_4= 1/(8\pi^4)$.
In absence of cubic nonlinearity management $\sigma=0$,
there exists an infinity of fixed points $a_c$ (for which $U_{av}'(a_c)=0$)
if the total mass is equal to the critical mass
$N_c= \pi \sqrt{3}/(2 \sqrt{\chi})$.
As it can be checked from (\ref{eq:odeave}) the Townes soliton is not stable
in the sense that it collapses if $N>N_c$ and it spreads out and vanishes
if $N<N_c$, so this theoretical solution of the quintic NLSE
with $N=N_c$ cannot be observed in practice.
In presence of strong cubic management $\sigma>0$,
we get the existence of a unique fixed point $a_c$ if $N >N_c$, with
\begin{equation}
\label{eq:ac}
a_c = \frac{\sqrt{2} \sigma N}{\pi \sqrt{(N/N_c)^2 -1}} \, \cdot
\end{equation}
The linear stability analysis shows that this fixed point is stable
and that the oscillation period is
\begin{equation}
\label{eq:tc}
T_c =  \frac{2 \sigma^2 N^2}{[ (N/N_c)^2 -1]^{3/2}} \, \cdot
\end{equation}
We are especially interested in solutions whose masses are just above the critical
mass, since our main goal is to prove that the Townes soliton
can be stabilized by cubic nonlinearity management.
In these conditions, the stable soliton width is rather large $\sim 1/(N-N_c)^{1/2}$,
and the soliton oscillation period is very long $\sim 1/(N-N_c)^{3/2}$.

\begin{figure}
\begin{center}
\begin{tabular}{cc}
{\bf a)}
\includegraphics[width=3.8cm]{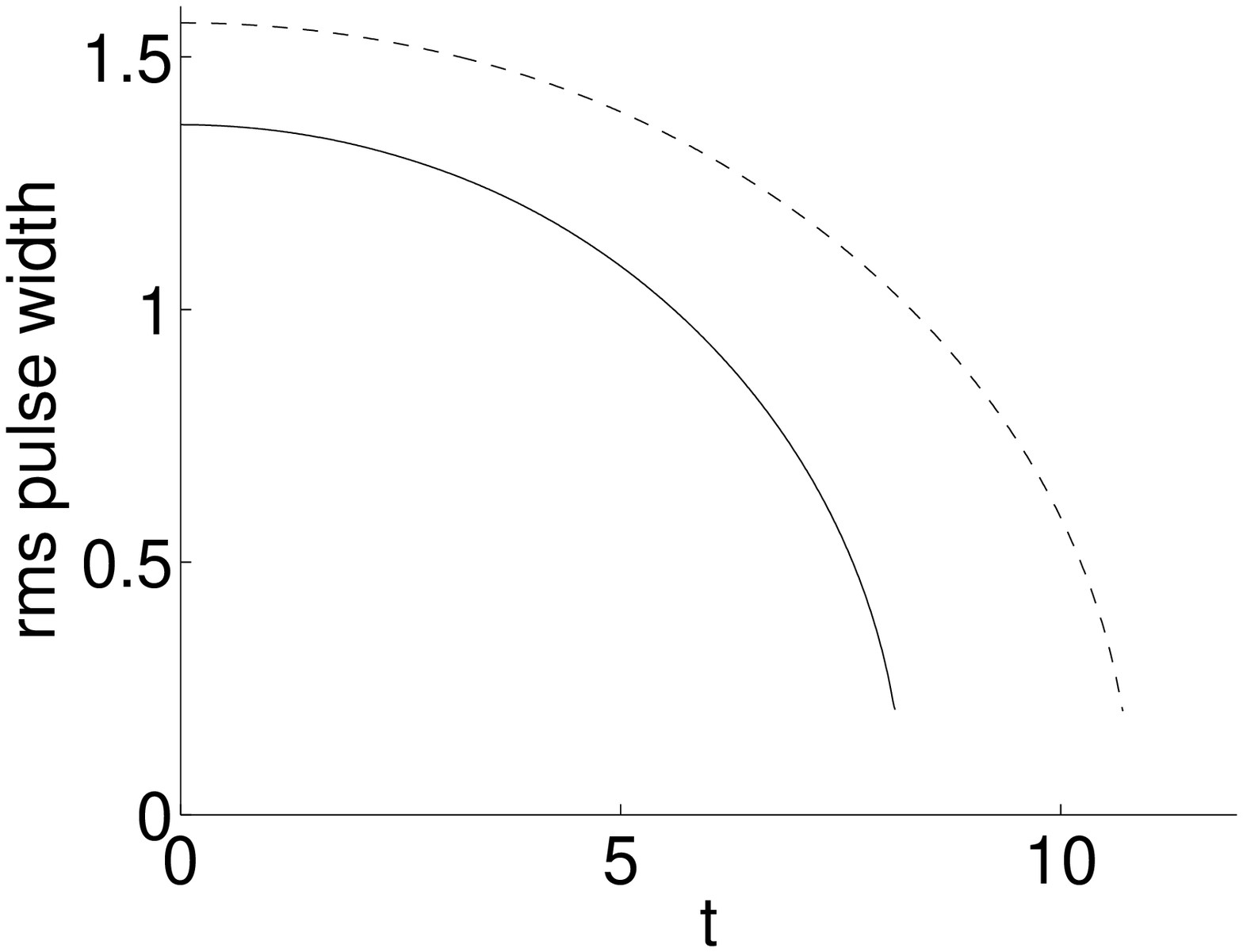}
&
{\bf b)}
\includegraphics[width=3.8cm]{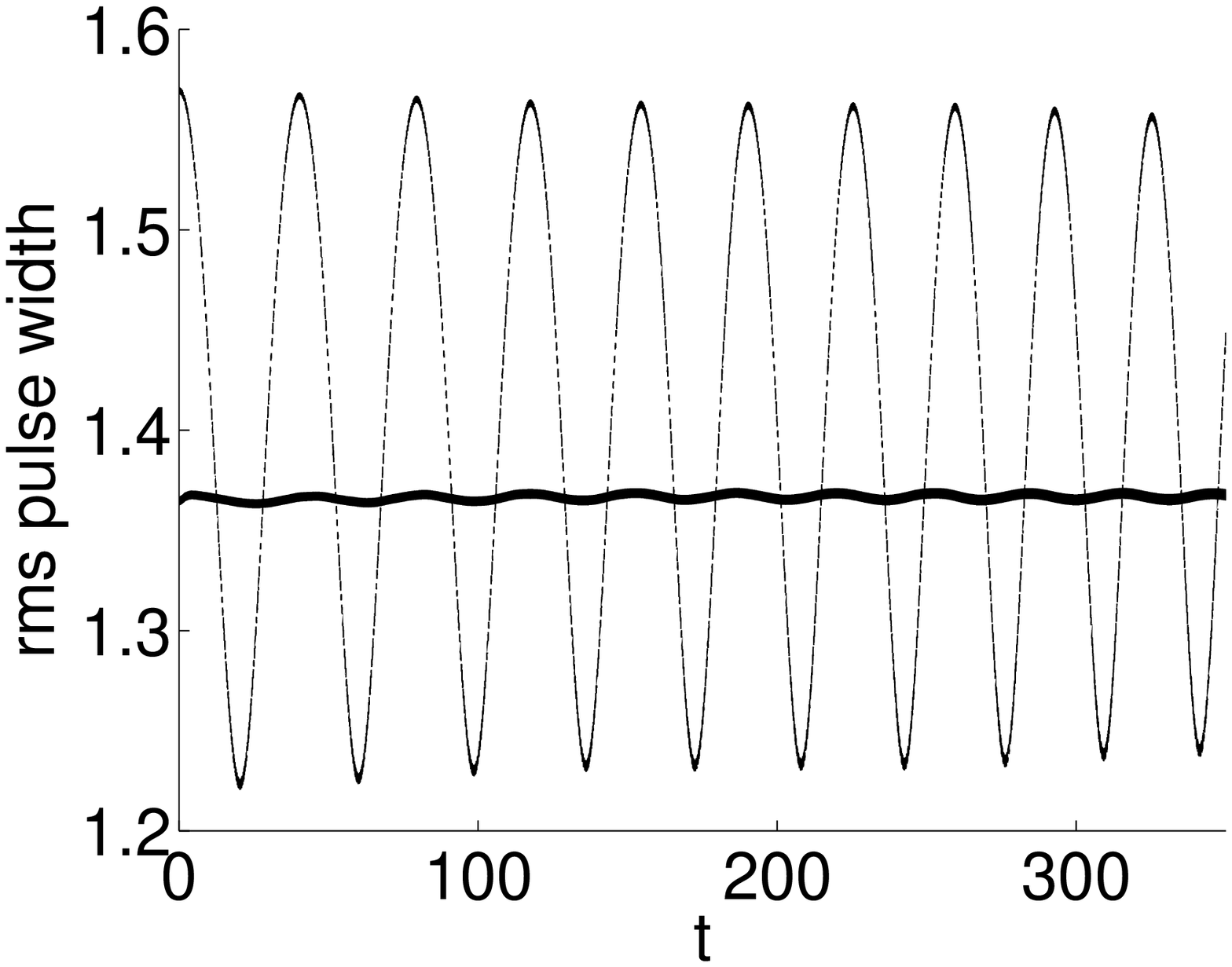}
\end{tabular}
\end{center}
\caption{Rms pulse width versus time. The initial pulse has a Townes shape
with mass $N=1.02 N_c\simeq 2.77$ and radius $a_0=a_c=0.88$
(corresponding to  an initial rms pulse width $a_{rms,0} =1.38$, solid line)
or $a_0=1.0$ ($a_{rms,0} =1.57$, dashed lines). Picture a:
collapse at finite time in absence of nonlinearity management.
Picture b: a periodic cubic nonlinearity management
$ \gamma(t)=10 \cos(50 t)$ is applied and the pulse rms width is plotted till time $t=350$.
\label{fig1}
}
\end{figure}

In the following numerical experiments, we take $\chi=1$,
so the critical mass is $N_c \simeq 2.72$,  and
we apply the management
 $\gamma(t)=10 \cos(50 t)$, which gives $\sigma=1/(5\sqrt{2})$.
We solve the CQ NLSE by a pseudo-spectral method
starting from a Townes soliton-shape  with a mass $N>N_c$ and radius $a_0$.
For $N=1.02N_c$ (Fig.~\ref{fig1}), the theoretical fixed point and oscillation
period are $a_c = 0.88$ and $T_c = 38$.
If we choose $a_0=a_c$ for the input pulse width,
then we observe in the numerical simulations
that the mean-square radius of the pulse is almost constant,
which shows that this solution is stable.
If we choose $a_0=1>a_c$, then we observe slow oscillations
around $a_c$ with a period very close to $T_c$.
These two observations demonstrate that the values $a_c$ and $T_c$
for the soliton parameters predicted by the variational analysis are
very accurate.

\begin{figure}
\begin{center}
\begin{tabular}{cc}
{\bf a)}
\includegraphics[width=3.9cm]{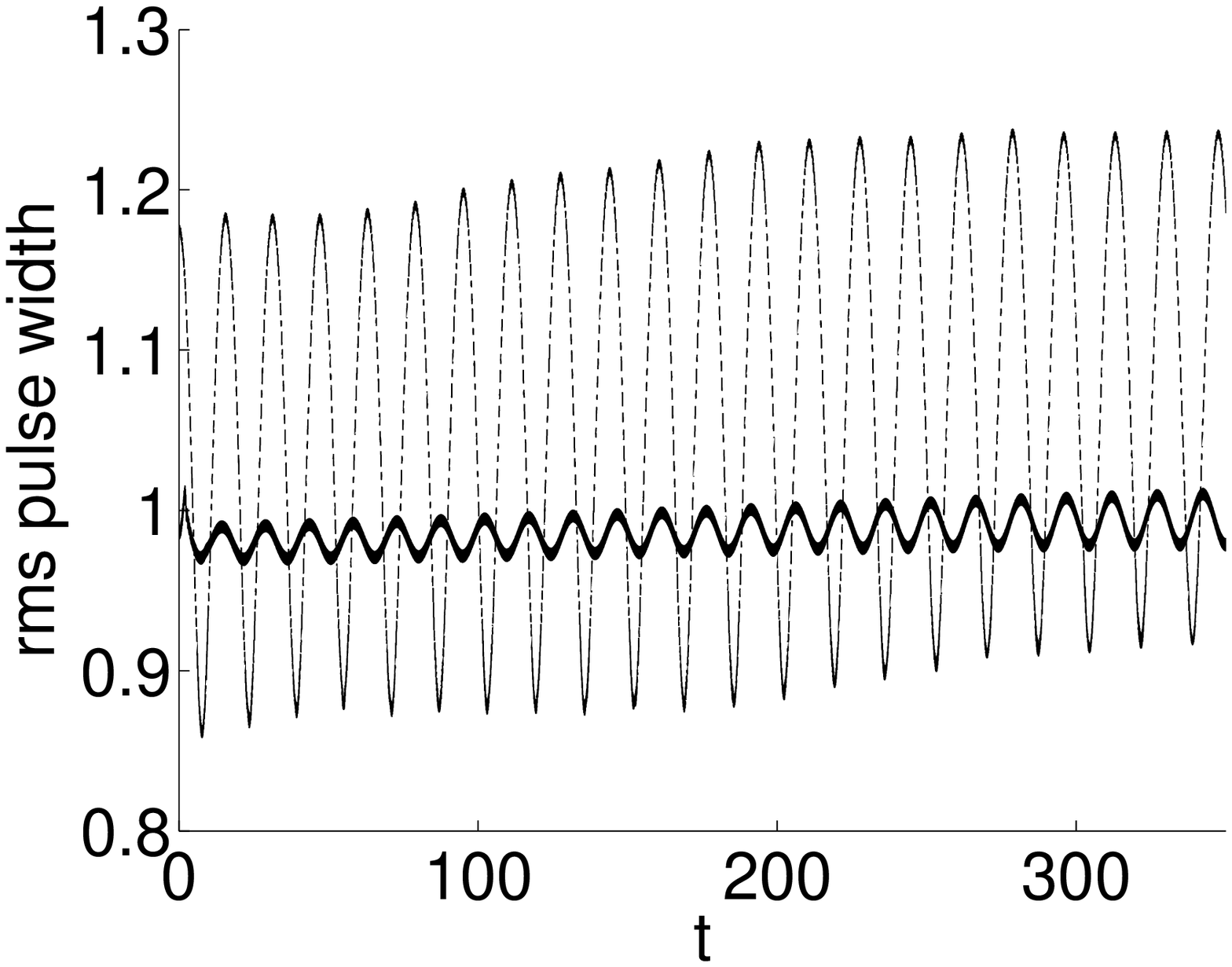}
&
{\bf b)}
\includegraphics[width=3.9cm]{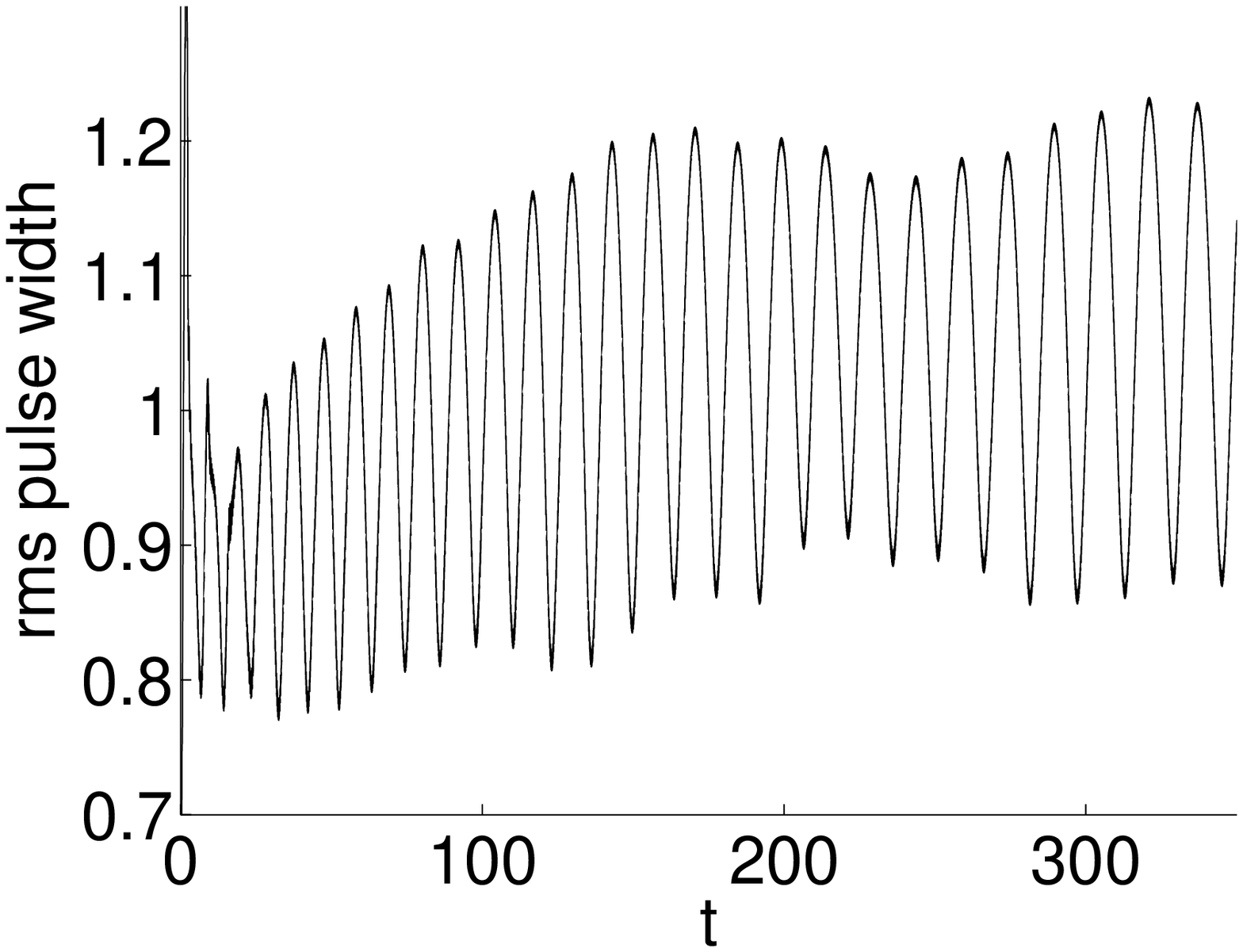}
\end{tabular}
\end{center}
\caption{Rms pulse width.
Picture a: $N=1.04 N_c$, $a_0 = 0.63$ ($a_{rms,0} = 0.99$, solid line),
$a_0 =0.75$ ($a_{rms,0} = 1.18$, dashed lines).
Picture b: $N=1.08 N_c$, $a_0 = 0.5$ ($a_{rms,0} = 0.79$, solid line).
\label{fig2}
}
\end{figure}

We have repeated the same experiments with different initial masses,
to check the validity of formulas (\ref{eq:ac}-\ref{eq:tc}).
For $N=1.04N_c$ (resp. 
$N=1.08N_c$),
the theoretical fixed points and oscillation
periods are $a_c = 0.63$ and $T_c = 14$ (resp.
$a_c = 0.42$ and $T_c = 5.1$).
As it can be seen in Fig.~\ref{fig2}b,
when the initial mass is significantly larger than the critical mass,
the pulse first experiences a strong distortion and emits radiation.
The resulting pulse is still stable, its mass is still well above the critical mass,
but the fixed point and oscillation period
are not well predicted  by the variational analysis.
If the initial mass is even larger, then this first phase leads to collapse, which shows that
the stabilizing effect of the cubic nonlinearity management $\gamma(t)=10 \cos(50t)$
is only efficient when the initial
mass is in the range $[N_c, 1.1 N_c]$.

\begin{figure}
\begin{center}
\begin{tabular}{cc}
{\bf a)}
\includegraphics[width=3.9cm]{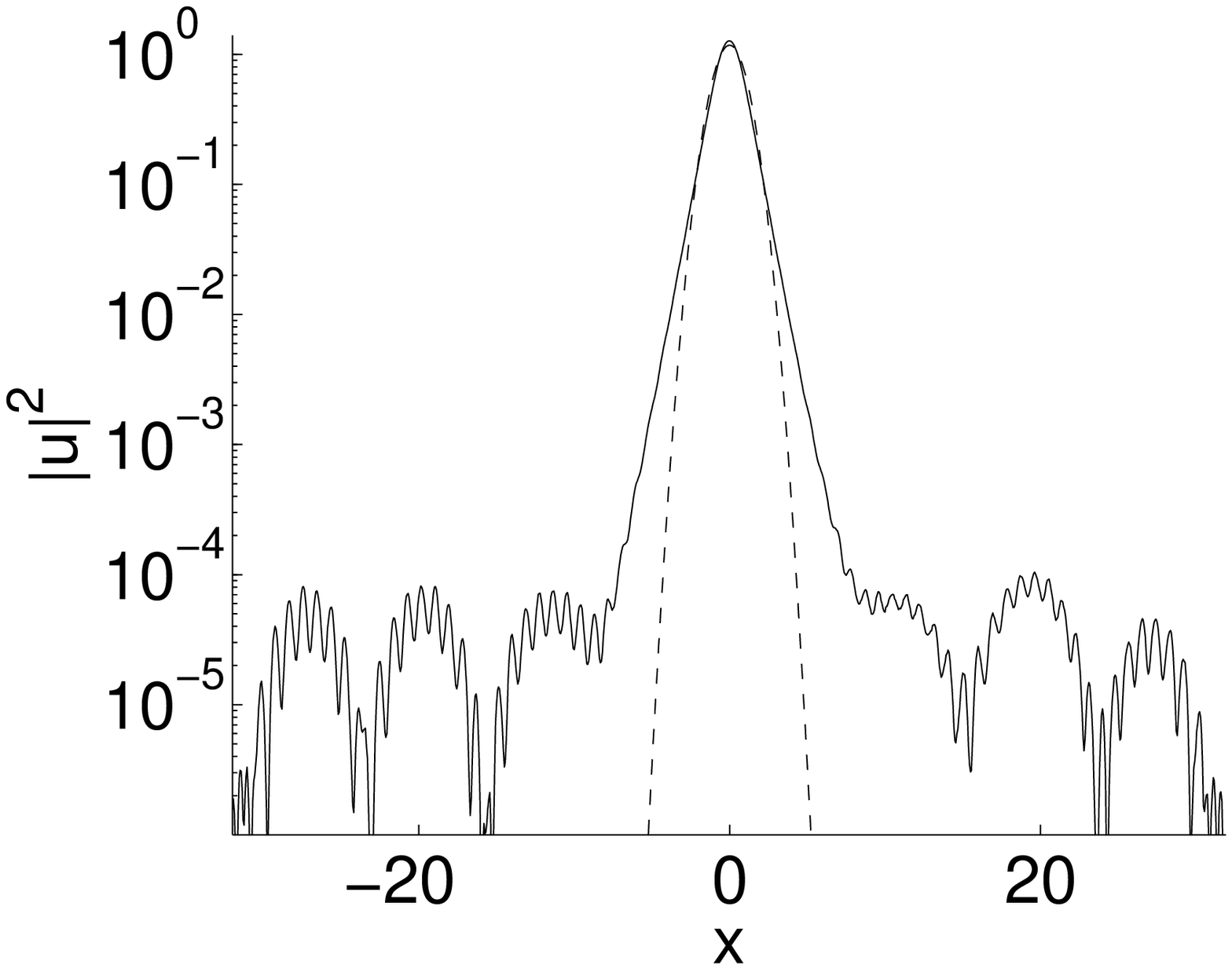}
&
{\bf b)}
\includegraphics[width=3.9cm]{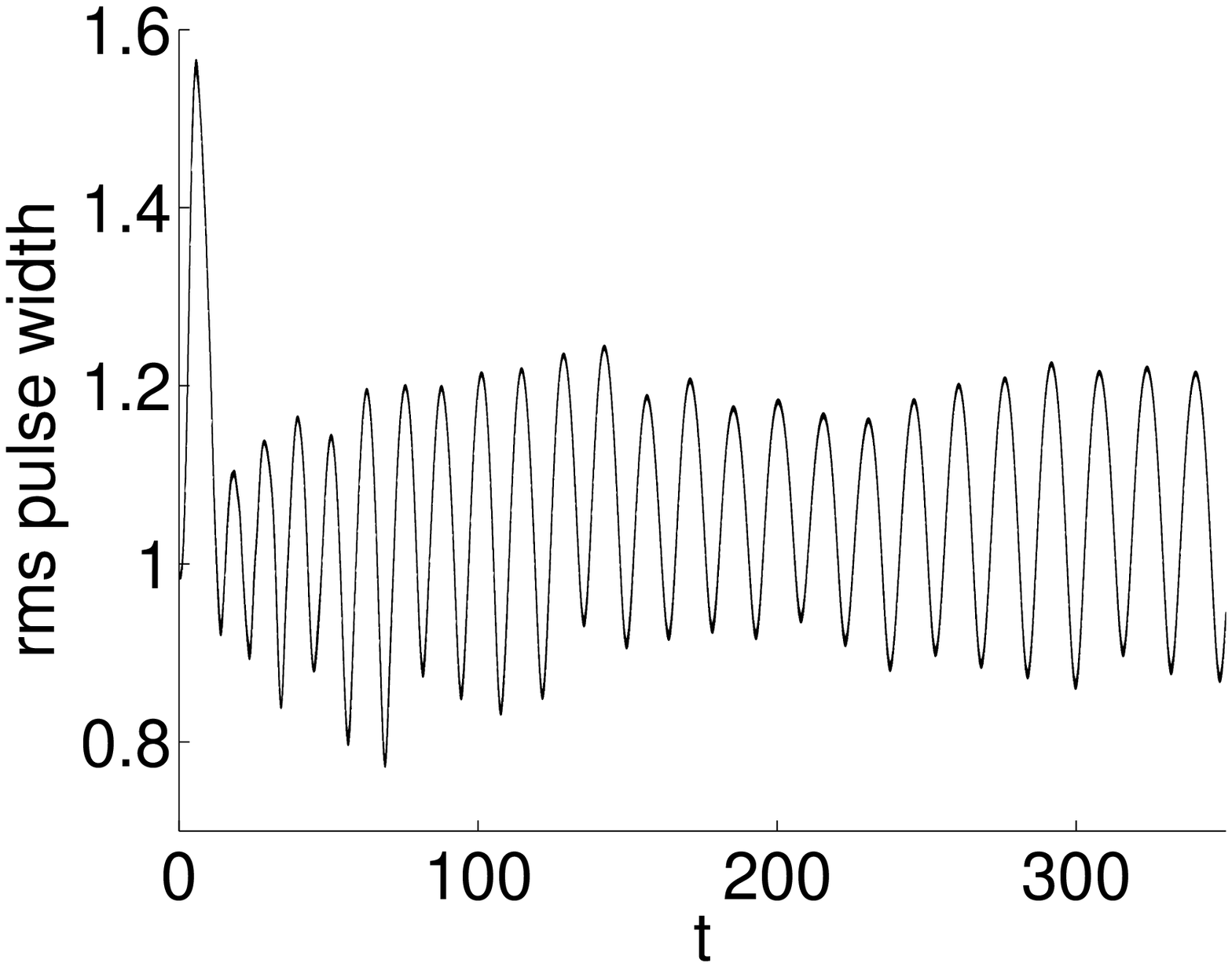}
\end{tabular}
\end{center}
\caption{Dynamics of a Gaussian pulse with initial mass $N=1.02 N_c = 2.92$
and radius $a_0=a_c = 1.4$ ($a_{rms,0}=0.99$).
Picture a: pulse profile at time $t=350$ (solid line)
compared to the input Gaussian profile (dashed lines).
Picture b: rms pulse width versus time.
\label{fig3}
}
\end{figure}

We now prove the robustness of the solution
of the CQ NLSE driven by cubic nonlinearity management
with respect to the initial pulse shape.
More exactly, we show that a stable solution can be obtained
with an initial pulse profile which is significantly different from the Townes profile.
The important conditions that have to be satisfied by the initial pulse is that its mass
should be just above the critical mass corresponding to the input pulse profile,
and that its initial radius should be chosen in the vicinity of the fixed point.
These conditions are imposed
by the analysis of Eq.~(\ref{eq:odeave}), and they are
confirmed by numerical simulations.
Let us consider an arbitrary pulse shape $Q$.
The critical mass is then $N_c = \sqrt{(3 D_1) / (D_3 \chi)}$
and the fixed point in presence of nonlinearity management is
$a_c = 2\sqrt{2} \sqrt{D_4/D1} N/\sqrt{(N/N_c)^2 -1}$,
with the period $T_c  = (4 \sqrt{2} \pi D_4/D_1^{3/2}) \sigma^2 N^2 /
[(N/N_c)^2-1]^{3/2}$.
A stable solution of the managed CQ NLS equation can be obtained by
injecting a pulse with shape $Q$ not too far from the Townes soliton,
with a mass just above $N_c$ and a radius close to $a_c$.
In the case of the  Gaussian ansatz $Q(x)=\exp(-x^2/2)$, we have
$D_1=1$,
$D_2=\sqrt{2/\pi}$, $D_3 = 2 /(\sqrt{3}\pi)$, and $D_4= 1/(3 \sqrt{3} \pi)$.
We report in Fig.~\ref{fig3} the results of  numerical simulations
carried out with $\chi=1$ so the critical mass is $N_c \simeq 2.86$.
We use the same management as above.
In a first phase ($t < 20$), the pulse emits radiation and its shape converges
to the one of the Townes soliton.
A similar phenomenon has been observed in numerical simulations carried out in \cite{Perez}
for the 2D cubic NLSE with cubic nonlinearity management.
After this transition period, the soliton width experiences
oscillations around the fixed point with an oscillation period close to the
theoretical value.

\begin{figure}
\begin{center}
\begin{tabular}{c}
\includegraphics[width=4.2cm]{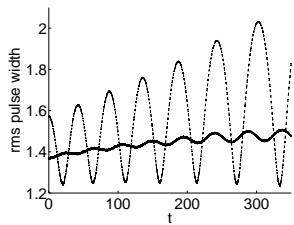}
\end{tabular}
\end{center}
\caption{Rms pulse width versus time. The initial pulse has a Townes shape
with mass $N=1.02 N_c\simeq 2.77$ and radius $a_0=a_c=0.88$
($a_{rms,0} =1.38$, solid line)
or $a_0=1.0$ ($a_{rms,0} = 1.57$, dashed lines). We apply a random
cubic nonlinearity management
$ \gamma(t)=10 m(50 t)$ where $m$ is zero-mean stationary
random process with a unit coherence time
compensated by a pinning scheme.
\label{fig4}
}
\end{figure}

We finally study the stabilization induced by random nonlinearity management.
Accordingly we now consider that $\gamma_1$ is a zero-mean stationary
random process. Considering Eqs.~(\ref{eq:odea}-\ref{eq:odeb}),
it can be seen that the important process is actually $\Gamma_1(t) =
\int_0^t \gamma_1(s) ds$, and it is critical that this process
does not grow in a diffusive manner, which would mean that cubic nonlinearity
accumulates. This condition is fulfilled
if $\gamma_1$ is the derivative of a stationary random process,
or if we apply a pinning scheme.
This technique was first introduced by \cite{Chertkov02} to compensate
for accumulated fiber dispersion. The periodic insertion of additional pieces of fiber with
well controlled lengths and dispersion values was found to prevent from pulse deterioration.
The pinning method can be applied to compensate for accumulated
cubic nonlinearity as well.
In Fig.~\ref{fig4} we show that random cubic nonlinearity management stabilizes
a Townes soliton in a manner similar to a periodic management.
However, we can detect a very slow spreading out, whose origin can be explained
by the imperfect compensation of the accumulated nonlinearity by the pinning scheme.

In conclusion we have analyzed the stabilizing role of the strong management of the
cubic nonlinearity in the 1D cubic-quintic NLSE. We have proved that the
averaged CQ NLSE, in a dramatic distinction from the non-modulated system,
supports stable solutions beyond the critical mass.
In particular the quintic Townes soliton, which is unstable in the non-modulated system,
becomes stable in presence of strong nonlinearity management if its width lies
in the vicinity of some fixed point.
This fixed point and the associated oscillation frequency are well predicted
by the variational approach applied to the averaged CQ NLSE.
We have also developed and applied a modified variational
approach to the CQ NLSE with strong management, which gives
the same values for the parameters of the stabilized soliton.
We have checked that a random management is also stabilizing, if we take care
to select a random process whose cumulative values are small.
These results can be applied for the search of nonlinearity managed solitons in double-doped optical fibers and BECs with three-body interactions.



\end{document}